\documentclass[utf8]{notfrontiersSCNS}

\usepackage{url,lineno,microtype,subcaption}
\usepackage[hidelinks]{hyperref}
\usepackage[onehalfspacing]{setspace}

\newcommand{\spinnthree}{SpiNN-3 }
\newcommand{\spinnfive}{SpiNN-5 }
\newcommand{\pynn}{PyNN }

\def\firstAuthorLast{Rowley {et~al.}} 
\def\Authors{Andrew G. D. Rowley$^{1*}$, Christian Brenninkmeijer${^1}$, Simon Davidson${^1}$, Donal Fellows${^1}$, Andrew Gait${^1}$, David R. Lester${^1}$, Luis A. Plana${^1}$, Oliver Rhodes${^1}$, Alan B. Stokes${^1}$, Steve B. Furber${^1}$}
\def\Address{$^{1}$Advanced Processor Technologies Group, School of Computer Science, University of Manchester, UK}

\begin{document}

\onecolumn

\firstpage{1}

\title[SpiNNTools: The Execution Engine for the SpiNNaker Platform]{SpiNNTools: The Execution Engine for the SpiNNaker Platform}
\author[\firstAuthorLast ]{\Authors} 

\address{} 

\correspondance{} 

\extraAuth{}

\title{SpiNNTools: The Execution Engine for the SpiNNaker Platform}
\author{Andrew G. D. Rowley$^{1*}$, Christian Brenninkmeijer${^1}$, Simon Davidson${^1}$,\\ Donal Fellows${^1}$, Andrew Gait${^1}$, David R. Lester${^1}$,\\ Luis A. Plana${^1}$, Oliver Rhodes${^1}$, Alan B. Stokes${^1}$, Steve B. Furber${^1}$}
\date{\today}

\maketitle

\begin{abstract}
Distributed systems are becoming more common place, as computers typically contain multiple computation processors. The SpiNNaker architecture \citep{Furber2013} is such a distributed architecture, containing millions of cores connected with a unique communication network, making it one of the largest neuromorphic computing platforms in the world \citep{Furber2016}.  Utilising these processors efficiently usually requires expert knowledge of the architecture to generate executable code.  This work introduces a set of tools (SpiNNTools) that can map computational work described as a graph in to executable code that runs on this novel machine.  The SpiNNaker architecture is highly scalable which in turn produces unique challenges in loading data, executing the mapped problem and the retrieval of data. In this paper we describe these challenges in detail and the solutions implemented.

\end{abstract}

\section{Introduction}
With Moore's Law \citep{Moore1965} coming to an end, one way to gain the advantage of more computing power is through parallel computing.  As such, distributed and parallel computing platforms are becoming more prolific in the real world. These range from computing clusters (e.g., Amazon Web Services \citep{Murty:2008:PAW:1407893}, the high throughput Condor platform \citep{condor-practice}), to crowd sourcing techniques (e.g., BOINC \citep{Anderson:2004:BSP:1032646.1033223}). Utilising these types of resources often requires expert knowledge to be able to create and debug code that is designed to be executed in a distributed and parallel fashion.  More recently, software stacks have been created which try to abstract this process away from the end user by the use of explicit interfaces \citep{Forum:1994:MMI:898758, Dagum:1998:OIA:615255.615542}, or defining the problem in a form which is easier to map into a distributed system \citep{Dean:2008:MSD:1327452.1327492}.

\begin{figure}
\centering
\includegraphics[width=10cm]{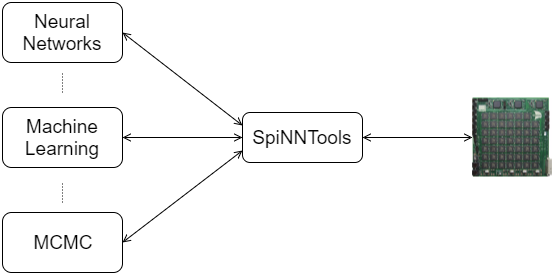}
\caption{Applications using SpiNNTools to control SpiNNaker}
\label{fig:high_level_tools}
\end{figure}

A SpiNNaker machine is one such distributed parallel computing platform; SpiNNaker is a highly scalable low power architecture in that each SpiNNaker board containing 48-chips can be connected with up to six other SpiNNaker boards. Each SpiNNaker chip contains 128 MiB of on-board memory, up to 18 ARM968E-S \citep{ARM968E} low-power 200 MHz processors, and a router for taking packets off and putting packets on to the communication fabric, allowing communication with six adjacent SpiNNaker chips simultaneously. Each chip uses up to 1W when all the processors are fully utilised, and chips and whole boards can be turned off when not in use to save energy. A growing number of users are now using SpiNNaker for various tasks, including Computational Neuroscience \citep{Albada2018} and Neuro-robotics (\citep{Richter2016, 10.1007/978-3-642-40728-4_59, 10.1007/978-3-319-12643-2_68}) for which the platform was originally designed, but also machine learning \citep{7280625}, and general parallel computation tasks, such as Markov chain Monte Carlo simulations \citep{7086903}.  The provision of a software stack for this platform aims to provide a base for the various applications, making it easier for them to exploit the full potential of the platform.  Additionally, users will gain the advantage of any improvement in the underlying tools without requiring changes to their software (or at most only minor changes should interface changes be required).  A basic overview of this approach is seen in Figure \ref{fig:high_level_tools}.

The SpiNNTools software stack\footnote{The software is open-source and available at \url{https://github.com/spinnakermanchester/}} to be detailed in this work allows the user to describe their computational requirements in the form of a graph, where the vertices represent the units of computation, and the edges represent communication of data between the computational units.  This graph is described in a high level language and the software then maps this directly onto an available SpiNNaker machine.  The SpiNNaker platform as a whole is intended to improve the overall execution time of the computational problems mapped on to it, and so the time taken to execute this mapping is critical; if it takes too long, it will dwarf the computational execution time of the problem itself.

This paper describes the functionality that this software provides as of version 4.0.0.  This description will proceed as follows.  Section \ref{sec:spinnaker_arch} describes the SpiNNaker architecture in more detail to explain the machine onto which to problems must be mapped.  We will then discuss the software with which the SpiNNaker application cores are programmed in section \ref{sec:spinnaker_core_software}, as well as look at previous tool chains for SpiNNaker that have tried to solve the same problem in the past in section \ref{sec:previous_software}.  This is followed by a discussion of the data structures required in section \ref{sec:data_structures}.  We then go in to the details of how these structures are used to map the graph on to the machine in section \ref{sec:tool_chain}, followed by an evaluation of the tools on some example applications that can be described as a graph in section \ref{sec:evaluation}.  We finally discuss further work to be done in section \ref{sec:future_work} and present conclusions in section \ref{sec:conclusions}.

\section{SpiNNaker Architecture}\label{sec:spinnaker_arch}

\begin{figure}
\centering
\includegraphics[width=10cm]{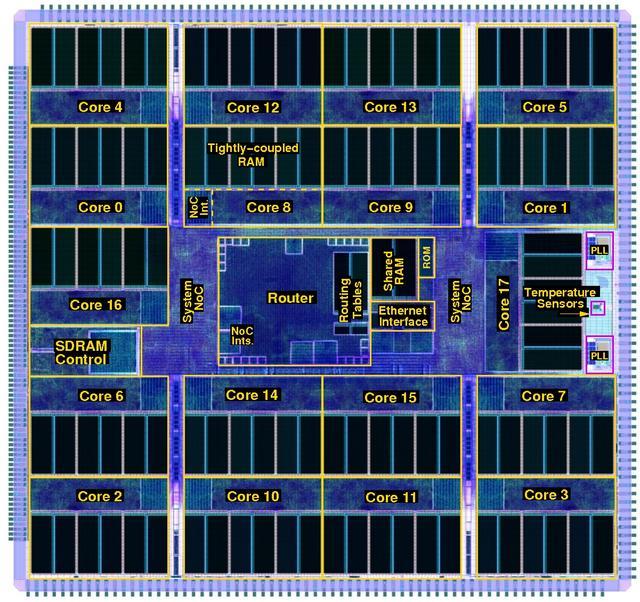}
\caption{The SpiNNaker chip in detail}
\label{fig:spinnaker_chip}
\end{figure}

In this section we will look at the SpiNNaker Architecture with respect to how the software will have to deal with the mapping of problems on to the machine.  A SpiNNaker machine is constructed from one or more SpiNNaker boards each of which houses a number of SpiNNaker chips. There are two production versions of the SpiNNaker board, known as \spinnthree and \spinnfive which have 4 and 48 chips respectively; the latter additionally has 3 FPGAs to allow it to be connected to up to 6 other boards to make up a larger SpiNNaker machine. Figure \ref{fig:spinnaker_chip} shows a graphical representation of a SpiNNaker chip, which shows that each SpiNNaker chip contains:
\begin{itemize}
\item up to 18 ARM968E-S \citep{ARM968E} processors (also referred to as cores), each of which has:
\begin{itemize}
\item 32 KiB of core-local, tightly-coupled instruction memory, referred to as ITCM, into which the entire processor scoped application code must fit,
\item 64 KiB of core-local, tightly-coupled data memory, referred to as DTCM, into which all locally scoped data and the application stack must fit.
\item A DMA controller for transferring data between the core-local and node-local memory.
\end{itemize}
\item 128 MiB of node-local, on-board memory in the form of SDRAM (not shown in the diagram as this is wire-bonded onto the chip), which can store large data structures, at the penalty of latency of access.
\item The SpiNNaker router.
\item On-board sensors for hardware monitoring.
\end{itemize}

The nature of the SpiNNaker chip has important implications for the software which can run on the system.  Firstly, it must be possible to break up the computation of the application into units small enough that the code for each part fits on a single core.  The SDRAM is shared between the cores on a single chip, and this property can be used by the application to allow cores to operate on the same data within the same chip.  A small amount of data can be shared with cores operating on other chips as well through communication via the SpiNNaker router.  The SpiNNaker boards can be connected together to form an even larger grid of chips, so appropriately parallelizable software could potentially be scaled to run on up to 1 million cores.

\begin{figure}
\centering
\includegraphics[width=10cm]{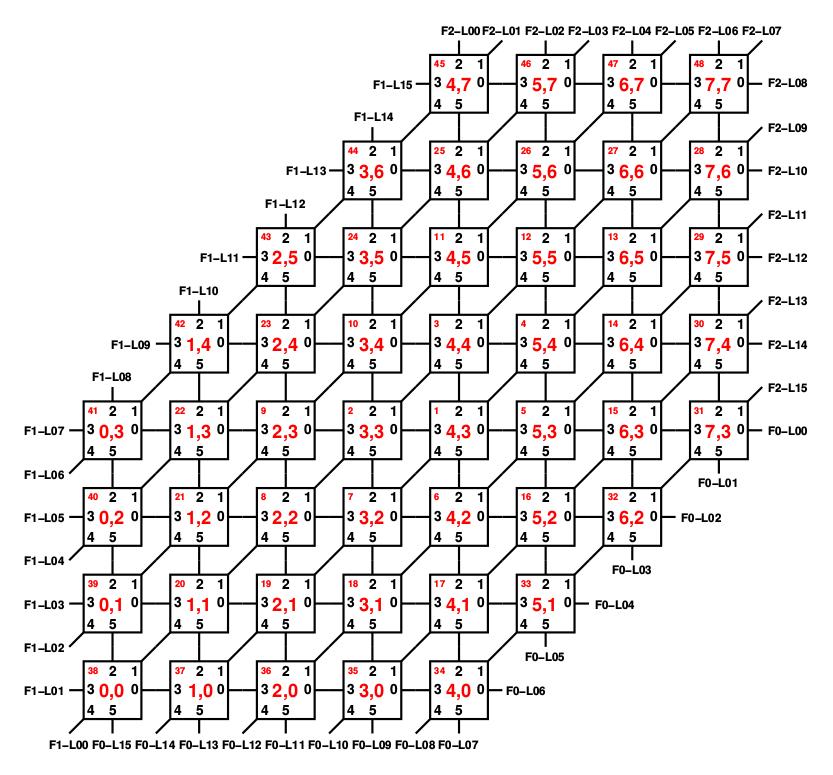}
\caption{SpiNNaker board wiring}
\label{fig:basic_48_chip_board_wiring}
\end{figure}

The SpiNNaker router on each chip has six links with which it sends and receives SpiNNaker packets \citep{Furber2013} from its neighbouring chips.  When arranged in a standard Cartesian 2D plane with the x-axis being labelled East/West and the y-axis being labelled North/South, the connections are usually made to chips in the directions of North, North East, East, South, South West and West. The router can also send and receive SpiNNaker packets to and from any of the processors running on the respective chip. The wiring of an individual \spinnfive board is shown in Figure \ref{fig:basic_48_chip_board_wiring}.

\begin{figure}
\centering
\includegraphics[width=15cm]{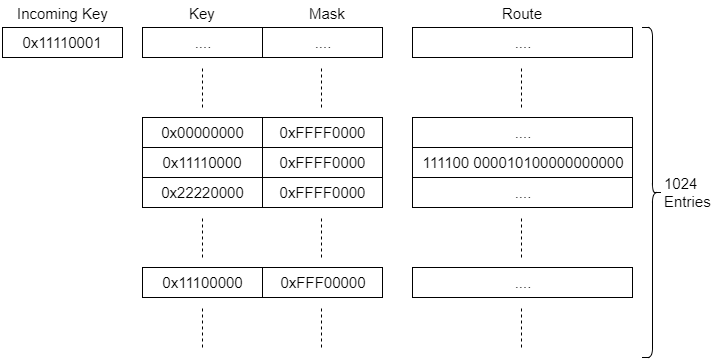}
\caption{The SpiNNaker Router TCAM.  An incoming key is matched against the entries in the table using the mask to determine which bits can be ignored and a route is determined, made up of 6 link bits and 18 processor bits, which indicates where the packet should be sent.  An entry might match with multiple keys once masked, in which case match which appears earliest in the table is used.}
\label{fig:router_cam}
\end{figure}

The SpiNNaker router is initially set up to handle the routing of system-level data.  The data to be sent by applications makes use of the multicast packet type, meaning that a packet sent from a single source can be routed to multiple destinations simultaneously.  Each of these packets consists of a key and optionally a payload.  To make multicast routing work, the routing tables of the router must be set up with an ordered list of up to 1024 entries, each of which has a key, a mask and a route, as shown in figure \ref{fig:router_cam}.  When a packet is received with a key which matches the key of an entry after a mask is applied, it is sent to the processor(s) and/or link(s) represented in the route of that entry.  The list is ordered so that the first such match is taken over any other.  If no match is found, the packet is routed out of the opposite link to the one on which it was received; thus packets will travel in a straight line through chips if not redirected by a routing entry.  If the packet was received from a local processor, and no entry matches, the packet is dropped.

Due to the asynchronous nature of the SpiNNaker machine's communication fabric, it is possible to cause deadlocks within the system by having loops of communication.  To avoid the traffic coming to a standstill, the SpiNNaker router can drop the next packet to be sent after a given time period (see \citep{Furber2013} for a more detailed explanation). If a packet is dropped, an interrupt is raised so that this can be detected and potentially dealt with (see later).

Each chip additionally has an Ethernet controller, although in practice only one chip is connected to the Ethernet connector on each board. The chip with the Ethernet connected to it is then called the Ethernet chip, and this is used to communicate with the outside world, allowing, for example, the loading of data and applications. Communications with other chips on a board from outside of the machine must therefore go via the Ethernet chip; system-level packets are used to effect this communication between chips. In practise, the Ethernet connector of every board in a SpiNNaker machine is connected and configured, though this isn't a requirement.

SpiNNaker machines are designed to be fault tolerant, so it is possible to have a functional machine with some missing parts. For example, it is normal that some of the SpiNNaker chips have 17 instead of 18 working cores, and sometimes even less than this as operational cores are tested more thoroughly than the testing done at manufacture. Additionally, machines can have whole chips that have been found to have faults, as well as some links missing between the chips and boards.  The machine includes memory on to which faults can be stored statically in a "blacklist", so that during the boot process these parts of the machine can be hidden to avoid using them.

SpiNNaker machines can be connected to external devices through either a SpiNNaker-Link connector, of which there is one on every 48-node board, or a SpiNN-Link SATA connector, of which there are 9 on each board; of those, 6 are used to connect to other boards.  This, along with the low power requirements, make the machine particularly useful for robotics applications, since the board can be connected directly to the robot without any need of other equipment.  The only requirement is that the external devices must be configured to talk to the machine using SpiNNaker packets.  The links can be configured to directly connect with the links connected to a subset of the SpiNNaker chips on the board, and so entries in the routing tables of the chip can be used to send packets to any connected device and similarly to route packets received from the devices across the SpiNNaker network.

\section{SpiNNaker Core Software}\label{sec:spinnaker_core_software}
This section discusses the software that runs on each core of the machine, and how this is programmed. The ARM-968 cores can execute instructions from the ITCM using the ARM or THUMB instruction sets; generally this code is generated from compiled C code using either the GNU gcc compiler\footnote{\url{https://developer.arm.com/open-source/gnu-toolchain/gnu-rm/downloads}} or the ARM armcc compiler\footnote{\url{https://developer.arm.com/products/software-development-tools/compilers/legacy-compiler-releases}}. To this end, a library known as SARK (Spinnaker Application Runtime Kernel), has been written which allows access to the features of the SpiNNaker core and chip \citep{Brown2015}. Additionally, software called SCAMP (Spinnaker Control And Monitor Program) has also been written which allows one of the cores to operate as a monitor processor through which the chip can be controlled \citep{Brown2015}, allowing, for example, the loading of compiled applications onto the other cores of the chip, the reading and writing of the SDRAM and the loading of the SpiNNaker routing tables. SCAMP software can also map parts of the machine known to be faulty out when it is first loaded. Thus when a description of the machine is obtained via SCAMP, only working parts should be present. The list of faults is stored on the boards themselves, and can be updated dynamically if other parts are found to be faulty in future.

The SCAMP code can be loaded on to one core on every chip of the machine, and these cores then communicate with each other allowing communication to any chip via any Ethernet connector on the machine. This communication makes use of the SpiNNaker Datagram Protocol (SDP) \citep{Furber2014}, which is encapsulated in to UDP packets for communication with external machines. Communication out of the machine from any core is achieved by using IP Tags. The SCAMP monitor processor on each Ethernet chip maintains a list of up to 8 IP Tags, which maps between values in the tag field of the SDP packets and an external IP address and port. When a packet is received that is destined to go out via the Ethernet (identified in the SDP packet header), this table is consulted and a UDP packet is formed containing the packet and this is sent to the IP address and port given in the table. The table can also contain Reverse IP Tags, where a UDP packet received from an external source is mapped from the UDP port in the packet to a specific chip and core on the machine, where the data of the packet is extracted and put in to an SDP packet before being forwarded to the given core.

In addition to the SARK library, another library called Spin1API has been written, which further abstracts the details of the SpiNNaker chip from the application \citep{Furber2014}. This allows each processor to run in a event driven manner, in the sense that there are events which generate interrupts and any application code can then be programmed to take advantage of these. A few of these events are as follows:
\begin{itemize}
\item Receiving packets of different types.
\item The expiration of an optionally periodic timer.
\item The completion of a DMA transfer between SDRAM and DTCM.
\end{itemize}

The problem of writing code to run on the cores of the SpiNNaker machine is discussed in more detail in \citep{Brown2015}, along with the types of applications which might be suitable to execute on the platform. In the rest of this work, it is assumed that the application has already been designed to run in parallel; the SpiNNTools software then works to map that parallel application on to the machine, execute it, and extract any results, along with any relevant data about the machine.  In addition, the software also includes some additional C code libraries which run on top of the Spin1API to provide interfaces for some of the more advanced features supported.

\section{Previous Software Versions}\label{sec:previous_software}
Using SpiNNaker machines in the past required end users to load compiled applications and routing tables manually onto the SpiNNaker machine through the use of the low level ybug software included with the aforementioned libraries\footnote{Available from \url{https://github.com/SpiNNakerManchester/spinnaker_tools/releases}}. Other software was then designed to ease the development of application code for end users. These consisted of: the aforementioned low level libraries SARK and Spin1API, and the monitor core software SCAMP; a collection of C code which represented models known in the neuroscience community and defined by the \pynn 0.6 language \citep{Davison2008} and  a collection of Python code which translates PyNN models onto a SpiNNaker machine.
These pieces of software were amalgamated into a software package known as PACMAN 48 \citep{Galluppi:2012:HCS:2212908.2212934} and supported the main end user community of computational neuroscientists for a number of years. These tools had the following limitations:
\begin{itemize}
\item Only supported SpiNNaker machines consisting of a single \spinnthree or \spinnfive board.
\item It was designed only to support the computational neuroscience community, and thus non-neural applications were not supported.
\item End users were still expected to have expertise in using the SpiNNaker hardware. To this end, they were expected to manually run separate scripts which together:
\begin{itemize}
\item Boot the SpiNNaker machine,
\item Load executables onto the SpiNNaker machine,
\item Load data objects onto SpiNNaker,
\item Check when the executing code finished,
\item Extract data from the SpiNNaker machine.
\end{itemize}
\end{itemize}

The intention of the SpiNNTools software stack is to support a range of suitable applications executing on the SpiNNaker hardware by providing a flexible abstraction layer where the end user represents their problem as a graph, which is then executed on the SpiNNaker machine without requiring such a low-level knowledge of how the machine works, thus overcoming the issues mentioned above.  This concept is briefly mentioned as ``The Uploader'' in \citep{Brown2015}, although, as will be demonstrated, the framework described herein is more complete in that it also: allows the user to express the generation of the data structures to be loaded (and possibly reloaded when changes have been made); controls the execution flow of the application where required; aids in the storage and retrieval of data recorded during the execution; and extracts and presents provenance data which can be used to determine the correctness of the results.

\section{Data Structures}\label{sec:data_structures}
\subsection{SpiNNaker Machines}
\begin{figure}
\centering
\includegraphics[width=15cm]{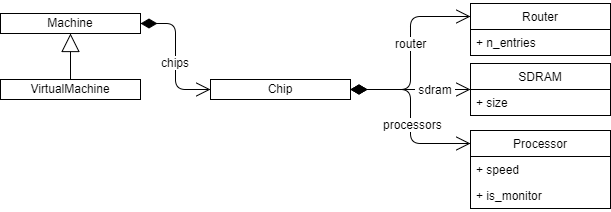}
\caption{The Python class hierarchy for SpiNNaker Machine representation.  The machine contains a list of chips, and each Chip contains a Router, SDRAM and a list of Processor objects, each with their respective properties.  A VirtualMachine can also be made, which contains the same objects but can be identified as being virtual by the rest of the tools.}
\label{fig:machine_uml}
\end{figure}

A SpiNNaker machine is represented as a set of Python classes as shown in figure \ref{fig:machine_uml}, with a main Machine class which then contains instances of classes for each of the parts of the machine represented. This data structure includes the important details of the machine for mapping purposes, including the chips, cores and links available, as well as the speed of each core, and the SDRAM available and the number of routing entries available on each chip (in case some of this resource is used by the system software, as it is in the case of SCAMP). As well as internally representing a physical, real-world machine with all its faults mapped out, this representation also allows the instantiation of a virtual machine for testing in the absence of connected hardware. The virtual machine can be further modified to simulate hardware faults and analyse the behaviour of the software.

The connection of external devices, such as a silicon retina or a motor, to the machine is represented using ``virtual chips''.  A virtual chip will be given coordinates of a chip that doesn't exist in the physical machine and is marked as virtual.  The coordinates don't have to align with the rest of the machine, as the location where the chip is connected to the other real chips in the machine is also identified.  This allows any algorithm to detect that virtual chips are present if necessary, and also to know where the connected real chip is to make use of that if needed.

\subsection{Graphs}
\begin{figure}
\centering
\includegraphics[width=15cm]{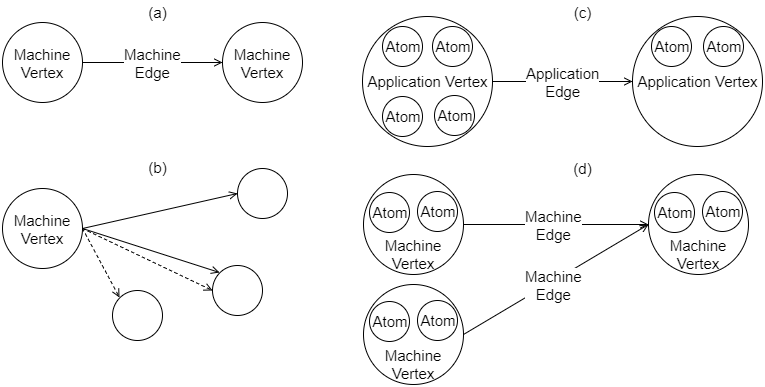}
\caption{Graphs in SpiNNTools. (a) A Machine Graph made up of two Machine Vertices connected by a Machine Edge, indicating a flow of data from the first to the second. (b) A Machine Vertex sends two different types of data to two subsets of destination vertices using two different Outgoing Edge Partitions, identified by solid and dashed lines respectively. (c) An Applications Graph made up of two Application Vertices, each of which contain two and four atoms respectively, connected by an Application Edge, indicating a flow of data from the first to the second. (d) A Machine Graph created from the Application Graph in (c) by splitting the first Application Vertex into two Machine Vertices which contain two atoms each.  The second Application Vertex has not been split.  Machine Edges have been added so that the flow of data between the vertices in still correct.}
\label{fig:graphs}
\end{figure}

A graph in SpiNNTools consists of vertices and directed edges between the vertices.  The vertex is considered to be a place where computation takes place and as such each vertex has a SpiNNaker executable binary associated with it. An edge represents some communication that will take place from a source, or pre-vertex to a target, or post-vertex.  An additional concept is that of the outgoing edge partition; this is a group, or partition, of edges that all start at the same pre-vertex, as shown in figure \ref{fig:graphs}(b).  This is useful to represent a multicast communication.  Note that not all edges that have the same pre-vertex have to be in the same outgoing edge partition; there can be more than one outgoing edge partition for each source vertex.  This represents different message types, which might be multicast to different sets of target vertices.  Thus each outgoing edge partition has an identifier, which can be used to identify the type of message to be multicast using that partition.

\begin{figure}
\centering
\includegraphics[width=\textwidth]{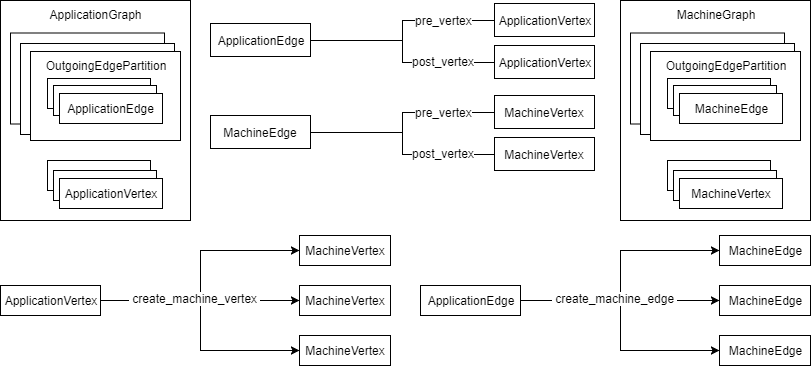}
\caption{The relationship between the graph objects.  An ApplicationGraph contains ApplicationVertex objects and OutgoingEdgePartition objects, which contain ApplicationEdge objects in turn.  A MachineGraph similarly contains MachineVertex objects and OutgoingEdgePartition objects, which contain MachineEdge objects in turn.  ApplicationEdge objects have pre- and post-vertex properties which are ApplicationVertex objects, and similarly MachineEdge objects and pre- and post-vertex properties which are MachineVertex objects.  An ApplicationVertex can create a number of MachineVertex objects for a subset of the atoms contained therein and an ApplicationEdge can create a number of MachineEdge for a subset of atoms in the pre- and post-vertices.}
\label{fig:graphs_uml}
\end{figure}

There are two types of graph represented as Python classes in the tools (a diagram can be seen in figure \ref{fig:graphs_uml}).  A Machine Graph, an example of which is shown in figure \ref{fig:graphs}(a) is one in which each vertex (known as a Machine Vertex) is guaranteed to be able to execute on a single SpiNNaker processor.  A Machine Edge therefore represents communication between cores.  In contrast, an Application Graph, an example of which is shown in figure \ref{fig:graphs}(c) is one where each vertex (known as an Application Vertex) contains atoms, where each atom represents an atomic unit of computation into which the application can be split; it may be possible to run multiple atoms of an Application Vertex on each core. Each edge (known as an Application Edge) represents communication of data between the groups of computational units; if one or more of the atoms in an Application Vertex communicates with one or more atoms in another Application Vertex, there must be an Application Edge between those Application Vertices. It is not guaranteed that all the atoms on an Application Vertex fit on a single core, so the instruction code for Application Vertices should know how to process a subset of the atoms, and how to handle a received message and direct it to the appropriate atom or atoms. The graph classes support adding and discovering vertices, edges and outgoing edge partitions.

As the vertices represent the application code that will run on a core, they have methods to communicate their resource requirements, in terms of the amount of DTCM and SDRAM required by the application, the number of CPU cycles used by the instructions of the application code in order to maintain any time constraints, and any IP Tags or Reverse IP Tags required by the application.  The Application Vertex provides a method that returns the resources required by a continuous range or slice of the atoms in the vertex; this is specific to the exact range of atoms, allowing different atoms of the vertex to require different resources.  The Application Vertex additionally defines the maximum number of atoms that the application code can execute at a maximum on each core of the machine (which might be unlimited), and also the total number of atoms that the vertex represents.  These allow the Application Vertex to be broken down into one or more Machine Vertices as seen in figure \ref{fig:graphs}(d); to this end, the Application Vertex class has a method for creating Machine Vertex objects for a continuous range of atoms.  A Machine Vertex can return the resources it requires in their entirety.

The graphs additionally support the concept of a Virtual Vertex.  This is a vertex that represents a device connected to a SpiNNaker machine.  The Virtual Vertex indicates which chip the device is physically connected to, allowing the tool chain to work with this to include the device in the network.  As with the other vertices, there is a version of the Virtual Vertex for each of the machine and application graphs.

\section{The SpiNNTools Tool Chain}\label{sec:tool_chain}
\begin{figure}
\centering
\includegraphics[width=15cm]{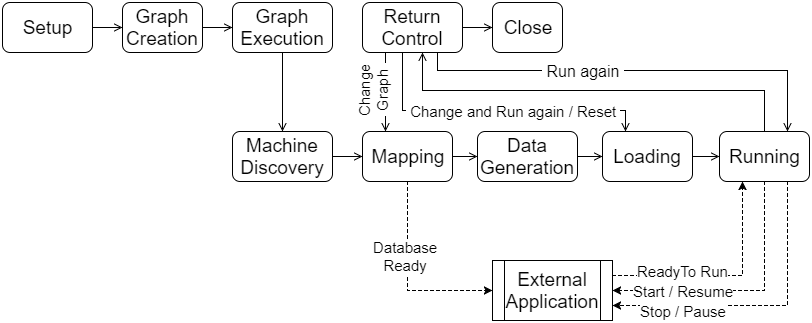}
\caption{The execution work flow of the tools in use within an Application.  Once control has returned to the application, the flow can be resumed at different stages depending on what has changed since the last execution.}
\label{fig:tools_workflow}
\end{figure}

The aim of the tool chain is to control the execution of a program described as a  graph on the SpiNNaker machine.  The software is executed in several steps as shown in figure \ref{fig:tools_workflow}, and detailed below.

\subsection{Setup}
The first step in using the tools is to initialise them.  At this point, the user can specify appropriate configuration parameters, such as the time step of the simulation, and the location where binary files can be located on the host machine.  The tools then set up the initially empty graphs, and read in configuration files for further options, such as the SpiNNaker machine to be used.  Options are separated out in this way to allow script-level parameters which might apply no matter where the script is run (like the timestep of the simulation), from user-level parameters, which will be different per-user, but likely to be common across multiple scripts for that user (like the SpiNNaker machine to be used).

\subsection{Graph Creation}
Once the tools have been initialised the user can add vertices and edges to either an application or machine graph.  It is an error to add vertices or edges to both of these structures.  The tool chain keeps track of the graph as it is built up.  Users can extend the vertex and edge classes to add additional information relevant to their own application.

\subsection{Graph Execution}
Once the user has built their graph, they then call one of the methods provided to start execution of the graph.  Methods are provided to run for a specified period of time, to run until a completion state is detected (such as all cores being in an exit state having completed some unit of work), or to run ``forever'' meaning that execution can be stopped through a separate call to the tools at some indeterminate time in the future, or the execution can be left on the machine to be stopped outside of the tools by resetting the machine.  The graph execution itself consists of several phases shown in the lower half of figure \ref{fig:tools_workflow} and detailed below.

\subsubsection{Machine Discovery}
The first phase of execution is the discovery of the machine to be executed on.  If the user has configured the tools to run on a single physical machine, this machine is contacted, and if necessary booted.  Communications with the machine then take place to discover the chips, cores and links available.  This builds up a python machine representation to be used in the rest of the tools.

If a machine is to be allocated, the tools must first work out how big a machine to request, by working out how many chips the user-specified graph requires.  If a machine graph has been provided, this can be used directly, since the number of cores is exactly the number of vertices in the graph.  The resources must still be queried, as the SDRAM requirements of the vertices might mean that not all of the cores on each chip can be used.  For example, a graph consisting of 10 machine vertices, each requiring 20 MB of SDRAM, and thus 200 MB of SDRAM overall, will not fit on a single chip in spite of their being enough cores.

If an application graph is provided, this must first be converted into a machine graph to determine the size of the machine.  This is done by executing some of the algorithms in the mapping phase (see below).

\subsubsection{Mapping}
The mapping phase takes the graph, and maps it on to the discovered machine.  This means that the vertices of the graph are assigned to cores on the machine, and edges of the graph are converted into communication paths though the machine.  Additionally, other resources required by the vertices are mapped on to machine resources to be used within the simulation.

If the graph is an application graph, it must first be converted to a machine graph.  This may have been done during the machine discovery phase as described previously.  To allow this, the algorithm(s) used in this ``graph partitioning'' process are kept separate from the rest of the mapping algorithms.

Once a machine graph is available, this is mapped to the machine through a series of phases.  This must generate several data structures to be used later in the process.  These include:
\begin{itemize}
\item A set of \textbf{placements} detailing which vertex is to be run on which core of the machine.
\item A set of \textbf{routing tables} detailing how communications over edges are to pass between the chips of the machine.
\item A set of \textbf{routing keys} detailing the range of keys that must be sent by each vertex in order to communicate over each outgoing edge partition starting at that vertex.
\item A set of \textbf{IP tags and reverse IP tags} identifies which external communications are to take place through which Ethernet-connected chip.
\end{itemize}

Note that once machine has been discovered, mapping can be performed entirely separately from the machine using the Python machine data structures created.  However, mapping could also make use of the machine itself by executing specially designed parallel mapping executables on the machine to speed up the execution.  The design of these executables is left as future work.

Mapping information can be stored in a database by the system.  This allows for external applications which interact with the running simulation to decode any live data received.  As shown in figure \ref{fig:tools_workflow}, the applications can register to be notified when the database is ready for reading, and can then notify the tools when they have completed any setup and are ready for the simulation to start, and when the simulation has finished.

\subsubsection{Data Generation}
The data generation phase creates a block of data to be loaded in to the SDRAM for each vertex.  This can be used to pass parameters from the python-described vertices to the application code to be executed on the machine.  This can make use of the mapping information above as appropriate; for example, the \textbf{routing keys} and \textbf{IP tags} allocated to the vertex can be passed to ensure that the correct keys and tags are used in transmission.  The graph itself could also be used to determine which \textbf{routing keys} are to be received by the vertex, and so set up appropriate actions to take upon receipt of these keys.

Some support for data generation and reading is provided by the tools at both the Python level where data can be generated in ``regions''; and at the C code level where library functions are provided to access these regions.  Other more basic data generation is also supported which simply writes to the SDRAM directly.

\subsubsection{Loading}
The loading phase takes all the mapping information and data generated, along with the application binaries associated with each machine vertex, and prepares the physical machine for execution.  This includes loading the \textbf{routing tables} generated on to each chip of the machine, loading the application data into the SDRAM of the machine, loading the \textbf{IP tags and reverse IP tags} into the Ethernet chips, and loading the application code to be executed.

\subsubsection{Running}
The running phase starts off the actual execution of the simulation, and if necessary, monitors the execution until complete.  Before execution, the tools wait for the completion of the setup of any external applications that have registered to read the mapping database.  These tools are then notified that the application is about to start, and when it is finished.

\begin{figure}
\centering
\includegraphics[width=10cm]{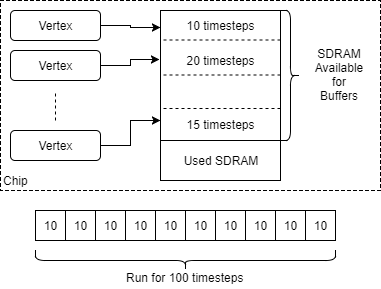}
\caption{Running vertices with recorded data.  The SDRAM remaining on each chip after it has been allocated for other things is divided up between the vertices on that chip.  Each is then asked for the number of time steps it can be run for before filling up the SDRAM.  The minimum number of time steps is taken over all chips and the total run time is split into smaller chunks, between which the recorded data is extracted and the buffer is cleared.}
\label{fig:running}
\end{figure}

Once a run is complete, application recorded data, and provenance data is extracted from the machine.  The provenance data includes:
\begin{itemize}
  \item Router statistics, including dropped multicast packets.
  \item Core-level execution statistics, including information on whether the core has kept up with timing requirements.
  \item Custom core-level statistics.  These depend on the application, but might include such things as the number of spikes sent in a neural simulation, or the number of times a certain condition has occured.
\end{itemize}
The log files from each core can also optionally be extracted.  During provenance extraction, each vertex can analyse the data and report any anomalies.  If the log files have been extracted, these can also be analysed and any ``error'' or ``warning'' lines can then be printed.

If a run is detected to have failed in some way, the tools will attempt to extract information about this failure.  A failure includes one of the cores going in to an error state, or if the tools have been run for a specific duration, if the cores are not in a completion state after this time has passed.  Log files will be automatically extracted here and analysed as previously discussed.  Any cores that are still alive will also be asked to stop and extract any provenance data so that this can also be analysed in an attempt to diagnose the cause of the error.

The run may be split into several sub-runs to allow for the limited SDRAM on the machine, as shown in figure \ref{fig:running}. After each run cycle, any recorded data is extracted from the SDRAM and stored on the host machine, after which the recording space is flushed, and the run cycle restarted.  This requires additional support within the binary of the vertex, to allow a message to be sent to the core to increase the run duration, and to reset the recording state.  This support is provided in the form of C code library functions to perform these operations, with callbacks to allow the user to perform additional tasks before resuming execution at each phase.  Additionally, the tools can be set up to extract and clear the core logs after each run cycle to ensure that the logs don't overflow.

The length of each run cycle can be determined automatically by the tools.  This is done by working out the SDRAM available on each chip after data generation has taken place.  This free space is then divided equally between the vertices on the chip, and each vertex is then asked how many units of time it can run for given that space.  The minimum such time is taken and used as the unit for each cycle; any left over time is then used up in the last run cycle.  To ensure that there is some space for recording, vertices can also reserve a minimum recording space.

At the end of each run phase, external applications are notified that the simulation has been paused, and are then notified again when the simulation resumes.  This allows them to keep in synchronization with the rest of the application.

\subsection{Return of Control / Extraction of Results}
Once the run cycles have completed, the tools return control to the executing script.  At this point, the user can interact with the graph again.  This includes the ability to extract any recorded data (see later), or make changes to the graph and/or the parameters before resuming the simulation.  The effect of any changes is detailed below.

\subsection{Resuming / Running Again}
The user can choose to resume the execution of the simulation, or to reset the simulation and start it again.  At this point, the tools must decide which of the aforementioned steps need to be run again.  If no changes have been made to the graph or the parameters, this can simply be considered an extension of the aforementioned ability for the tools to run the code in phases.  The minimum time calculated previously is respected again here and the tools will then run in cycles of this unit of time.  Note that this means that if the first run-time is shorter than that required to fill the remaining SDRAM space (and thus only one run cycle was required previously), this time is taken as the minimum.  This is because the buffers will have already been initialised to record for this amount of time.  An extension to this work then is to allow the buffers to be sized to use up all of the remaining SDRAM regardless of the run time, and then allow runs in units of less than or equal to the time that uses all of this space.

If the parameters of any of the vertices or edges have been changed, the vertex can be set up to allow the reloading of these changes.  It is expected that this can be supported where the change won't increase the size of the data, and so can overwrite the existing data, such as a change in neuron state update parameters in a neural network.  Any increase in the size of the data, such as an increase in the number of synapses in a neural network, would likely require a remapping of the graph on to the machine as the SDRAM is likely to be packed in such a way as to not allow the expansion of the data for a single core; it is left to the vertex to make this decision however.

Any change to the graph, such as the addition of a vertex or edge, is likely to require that the mapping phase take place again.  This may even result in a new machine being required should the size of the graph increase to this degree.  This will mean that all the other phases will also have to be executed again.

\subsection{Closing}
Once the user has finished simulating and extracted any data, they can tell the tools that they are finished with the machine by closing the tools. At this point, the tools reset and release any machines that have been reserved, and so recorded data will no longer be available.  If the tools were told to run the network for an indeterminate length, this would also result in the extraction and evaluation of any provenance data at this stage.

\subsection{Algorithms and Execution}
In order to run each of the above phases, the tools executes a series of algorithms.  The algorithms consume various inputs that are made available by the tools and by other algorithms, and produce various outputs.  These inputs and outputs are not constrained in any other way; thus algorithms are not constrained to produce only one output.  This could be useful in, for example, mapping, where an algorithm could be made to produce both placements and routing tables which have been optimised together.  This is in contrast to restricting the algorithms to specific tasks, where the output might then be less optimal, such as having a specific algorithm for generating placement and another for generating routing tables.

\begin{figure}
\centering
\centering
\includegraphics[width=12cm]{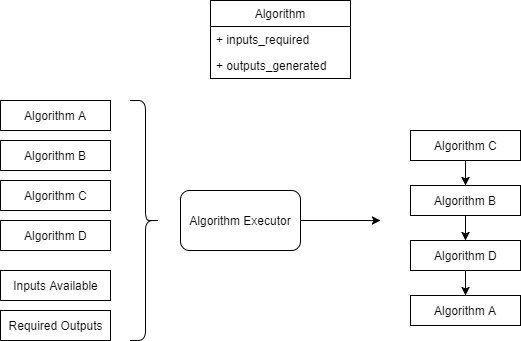}
\caption{Algorithms being run by the algorithm execution engine.  The executor is provided with a list of algorithms to run, a set of input items and a set of output items to produce.  It then produces a workflow for the algorithms accounting for their inputs required and outputs produced.}
\label{fig:algorithm_executor}
\end{figure}

To support this form of execution, the tools implement a workflow execution system, shown in figure \ref{fig:algorithm_executor}.  This examines the algorithms to be run in terms of the inputs required and outputs generated in order to compute an execution order for the algorithms.  Input and output ``tokens'' are also supported; these indicate implicit inputs and outputs; for example, a token might be used to represent that data has been loaded on to the machine, thus an algorithm can generate this as an output, and another can require that this has been completed before execution.

The algorithms themselves are not discussed here in detail other than those mentioned above.  A more detailed discussion of the mapping algorithms is discussed in \citep{Heathcote2016}.  The tools also include algorithms for routing table compression, which is discussed in \citep{Mundy2016a}.  Many of the other algorithms are currently simplistic in nature; these can be replaced in the future should other algorithms be found to perform more efficiently and / or effectively.

\subsection{Data Recording and Extraction}
\begin{figure}
\centering
\includegraphics[width=17cm]{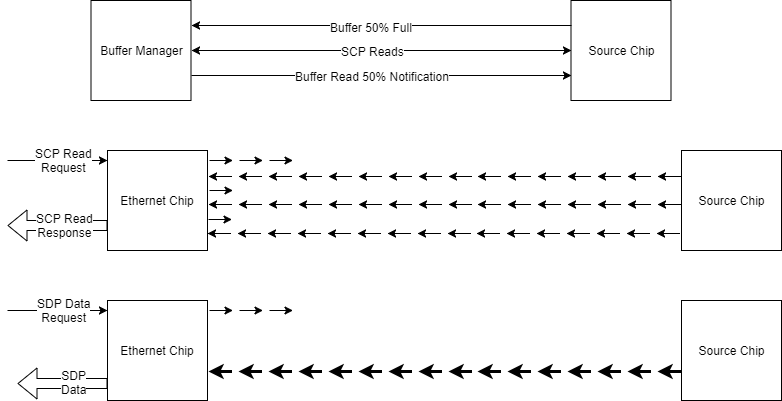}
\caption{Data Buffering and Extraction.  \textbf{Top:} The buffer manager is used to read back recorded data during execution; when the buffer contains some data, the buffer manager is notified and attempts to read the data, notifying the data source once this has been done to allow the space to be reused.  \textbf{Middle:} Data reading done using SCAMP; each read of up to 256 bytes is further broken down in to a number of request and read cycles on the machine itself, where the packets used contain only 24-bits of data each.  \textbf{Bottom:} Data reading done using multicast messages; the initial request is all that is required, after which the data is streamed using packets containing 64-bits of data.  The machine is set up so that these packets are guaranteed to arrive, so no confirmation is required.}
\label{fig:data_recording_extraction}
\end{figure}

As mentioned previously, the tools support the recording of data in such a way as to cope with the limited nature of the SDRAM on the machine.  To support this, a ``buffer manager'' is provided, which is used to keep track of and store the buffers of data as they are extracted from the machine.  This can additionally support the live extraction of buffers whilst the simulation is running, as shown in figure \ref{fig:data_recording_extraction}(Top); cores configured with the provided library can contact the host machine when the recording space is getting full and the tools can then attempt to extract the data.  In general, the bandwidth of the Ethernet of the machine is not fast enough for this to be effective, and data tends to be lost.

The SCAMP software supports the reading of SDRAM through SDP messages.  This works through a request and response system, where each SDP message can request the reading of up to 256 bytes of data.  Additionally, to transmit the SDP message to chips which are not connected to the Ethernet, this message must be broken down in to SpiNNaker network messages, and then reconstructed on receipt; an overview of how this process works is shown in figure \ref{fig:data_recording_extraction}(Middle).  This results in speeds of around 8Mb/s when reading from the Ethernet chip, and around 2Mb/s when reading from other chips.

In order to speed up the extraction of data, the tools include the ability to circumvent this process, an overview of which is shown in figure \ref{fig:data_recording_extraction}(Bottom).  To facilitate this, firstly the machine is configured so that packets can be sent with a guarantee that none of them are ever dropped; this can be done in this scenario because exactly one path through the machine will be used by each read, so deadlocks cannot occur.  Next, one of the cores on each chip is loaded with an application that can read from SDRAM and stream multicast messages to another application loaded on to a core on the Ethernet chip, which then forms these into SDP messages to be streamed to the host along with a sequence number in each SDP packet.  The host then gathers the SDP packets, and notes which sequences are missing.  The missing sequences are then requested again from the machine; this is repeated until all sequences have been received.  This has numerous advantages over the SDP request-and-response mechanism: the SDP is only formed at the Ethernet chip, thus the headers don't get transmitted across the SpiNNaker fabric; and the host only sends in a single request for data and then a single request for each group of missing sequences, and thus doesn't have to wait for each chunk of 256 bytes between sending requests.  This results in speeds of up to 40Mb/s when reading from any chip on the machine; there is no penalty for reading from a non-Ethernet chip.

Once this protocol was implemented, we discovered that the Python code had trouble keeping up with the speed at which the data was received from the machine.  We therefore implemented a version of the data reception in C++ and Java that could interface with the Python code.  This then allows the use of the Ethernet connection on multiple boards simultaneously, allowing the data extraction speed to scale with the number of boards required for the simulation, up to the bandwidth of the network connected to the host machine.

\subsection{Live Interaction}
We have previously mentioned that external applications can interact with a live simulation, making use of the mapping database.  Additional support for this interaction is provided by the tools.  This support is split into live data output and live data input.

\begin{figure}
\centering
\includegraphics[width=12cm]{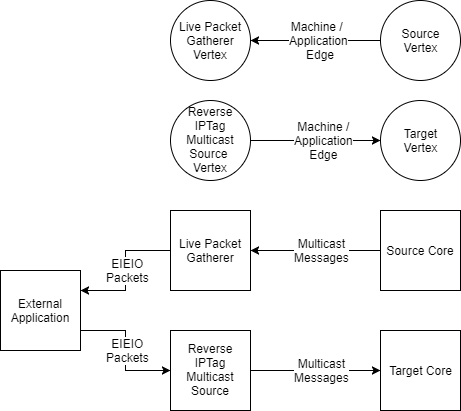}
\caption{Live Interaction with Vertices.  \textbf{Top:} To indicate that live output is required, an edge is added from the vertex which is the source of the data to the Live Packet Gatherer vertex in the graph.  To indicate the live input is required an edge is added from the Reverse IP Tag Multicast Source vertex to the target of the data in the graph.  \textbf{Bottom:} The effect of adding the edges to the graph is that multicast messages will be sent from the core (or cores) of the source vertex to the core running the Live Packet Gatherer, which will then wrap the messages in EIEIO packets forward them to a listening external application; and EIEIO packets received from an external application will be decoded by the Reverse IP Tag Multicast Source core and sent onwards as multicast messages to the target core (or cores).}
\label{fig:live_io}
\end{figure}

Live data output support is performed by a vertex called the ``Live Packet Gatherer'', which will package up any multicast packets it receives and send them as UDP packets using the EIEIO protocol \citep{10.1007/978-3-319-26561-2_79}.  It is configured by adding edges to the graph from vertices that wish to output their data in this way.  This has the advantage of being able to tap into the existing multicast streams that are already being used to communicate within the machine; this same data can be sent out of the machine by the simple addition of an edge to the graph, as shown in figure \ref{fig:live_io}.

Live data input support is provided via a vertex called the ``Reverse IP Tag Multicast Source'', which will unpack and send multicast packets using the same EIEIO protocol.  As with the Live Packet Gatherer, this vertex can then be configured by simply adding edges from it to the vertices which are to receive the messages.

External applications that would like to make use of this support can read the mapping database to determine the multicast keys to be received in the case of live output, or to be sent in the case of live input.  Support for this interaction is provided in the tools in both Python code and host-based C++ code.

\subsection{Dropped Packet Re-injection}
As mentioned in section \ref{sec:spinnaker_arch}, when a packet is dropped an interrupt is raised allowing a core to detect and capture the dropped packet.  The tools include software that runs on the SpiNNaker machine to detect this interrupt and then capture the packets that have been dropped.  These are stored until a time at which the router is no longer blocked, and so can safely send the packet onwards.  This helps in those applications where the reliable transmission of packets is critical to their operation.

There is only one register within the SpiNNaker hardware to hold a dropped packet. If a second packet is dropped, this packet will be completely unrecoverable; an additional flag is set in this scenario so the re-injection core can detect this and count such occurrences.  This count is reported to the user at the end of the execution so that they know that something may not be correct in their simulation results.

\section{Use Cases}\label{sec:evaluation}
In this section we will look at two example applications that can be described as a graph, and that work well with the SpiNNaker machine.  These applications will be used to show how the tools work to support the use of the machine.

\subsection{Conway's Game of Life}
\begin{figure}
\centering
\includegraphics[width=10cm]{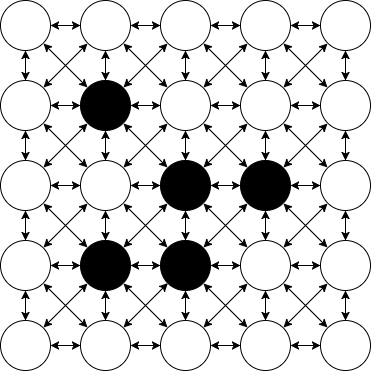}
\caption{Conway's Game Of Life on a 5x5 grid as a Machine Graph.  Every Machine Vertex is connected to each of it's 8 neighbours bi-directionally; this requires two Machine Edges for each bi-directional connection.  The initial state of each Vertex is either alive (black) or dead (white).}
\label{fig:conways}
\end{figure}

Conway's Game of Life \citep{gardner1970a} consists of a collection of cells which are either alive or dead based on the state of their neighbouring cells. A diagram of an example Machine Graph of this problem is shown in figure \ref{fig:conways}. The vertices of the graph of this application are each a cell in the game; given the state of the surrounding cells this cell can compute whether it is dead or alive in each step, and then send that to its neighbours.  It similarly receives the state of the neighbours as they are transmitted and again uses this to update its own state. The edges of the graph are thus between adjacent cells in a grid, where each vertex is connected bidirectionally to its eight surrounding neighbours. The game proceeds in synchronous phases, where the state of cells in a given phase are all considered at the same time.

Graphs of this form are highly scalable on the SpiNNaker system, since the computation to be performed at each node is fixed, and the communication forms a regular pattern which does not increase as the size of the board grows. Thus once working, it is likely that any size of game can be built, up to the size of the available machine. This type of graph would also likely be suited to finite element analysis \citep{Finite_Element_Analysis_Applications_2018} problems, provided that the data to be transmitted can be broken down into SpiNNaker packets. This problem thus works well as an archetype.

It will be assumed that we have built the application code which will update the cell based on the state of the surrounding cells.  This will update the state once per time step of the simulation based on the received state from the surrounding cells, and then send its own new state out using the given key.  It can also record its state at each time step in the simulation.  The set up of this application is as follows:
\begin{itemize}
  \item A Conway vertex is created which extends the machine vertex class.
  \item A number of Conway vertices are added to the graph to make up the board.  These are stored in such a way that finding an adjacent vertex in the grid is easy.
  \item A machine edge is added between each pair of adjacent vertices, in each direction.
  \item Each machine vertex generates data for the vertex, which includes the key to be sent by that vertex and the number of time steps to run for.
  \item Each machine vertex can tell the tools how many time steps it can run for given an amount of SDRAM available for recording.
  \item Each machine vertex contains code to read the state that is recorded at each time step using the Buffer Manager.
\end{itemize}

Once the graph is built, the script starts the execution of the graph. During this execution, the tools will obtain a machine description and use this with the machine graph to work out a placement of each of the vertices and a routing of the edges between these placements, along with an allocated key for each of the vertices.  The tools will then ask each vertex how many time steps it can record for based on the available SDRAM after placement is complete, and the resources used on each chip can therefore be determined.  Each vertex will then be asked to generate its data based on the mapping and timing information.  The tools will then load the generated data on to the machine along with the routing tables and application code, and start the execution of the cores.  It will wait an appropriate amount of time for the cores to stop, and then check their status.  Assuming this is successful, control will return to the script. This can then request the recorded states from each of the vertices and display this data in an appropriate way.

A future version could have a Conway vertex that can have multiple cells within each machine vertex, which would then allow for an application vertex of cells.  This would have a single large application vertex which would represent the whole game board and an application edge for each of the 8 directions of connectivity, each in its own Outgoing Edge Partition to indicate that different keys are required for each of the directions.  This would require that the vertex would have to cope with the reception of multiple neighbour states, which would make the application code itself more complex; for example, it would have to cope with multiple incoming keys from each direction, each of which would target a different cell within the grid.

Another possible extension to this application is to extract the state during execution and display this as the application progresses.  This would require the addition of the Live Packet Gatherer vertex (described above) to the graph, and an edge from each of the Conway vertices to this vertex.  The script would then indicate, before executing the graph, that there is an external application that would like to receive the data.  This application will receive a message when the mapping database has been written, at which point, it can set up a mapping between multicast keys received and positions in the game board, responding when it has completed its own setup.  The tools will then notify this application that the simulation is starting, and the application will then receive the same state messages as the vertices receive, which it can use to update the display of the game board.

\subsection{Spiking Neural Networks}
\begin{figure}
\centering
\includegraphics[width=15cm]{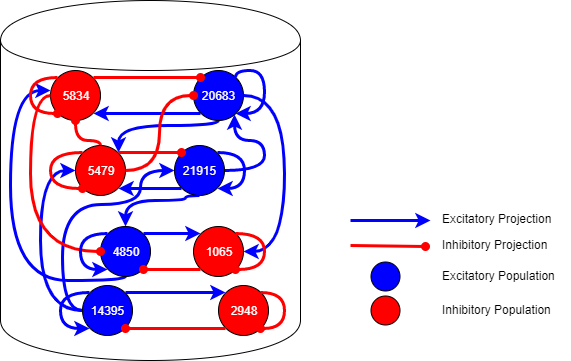}
\caption{A neural network topology of a $1mm^{2}$ area of Microcortial Column found within the mammalian brain.}
\label{fig:microcortical_column}
\end{figure}

The SpiNNaker machine is primarily designed to simulate spiking neural networks \citep{Furber2013}.  In particular, we consider the simulation of a micro-cortical column found within mammalian brains; that is, a model of the neurons within a structure underneath a $1mm^{2}$ area of the surface of the generic early sensory cortex \citep{Potjans2014}. Figure \ref{fig:microcortical_column} shows the of the groups of neurons (Populations) in this network and connectivity between them (Projections). In a spiking neural network, the vertices are groups of point neurons (as a single core can simulate more than one neuron); the computation required is the update of the neuron state in response to spikes received from connected neurons. The edges are then groups of synapses between the neurons, over which spikes are transmitted. These are potentially unidirectional, and are likely to be more heterogeneous in nature than the regular grid pattern seen in Conway's Game of Life.

The problem of spiking neural networks is clearly well suited to the architecture, as this is what it was designed for, but the heterogeneity of the network, and the fact that multiple neurons are computed on each core means that some networks will be more suited to the platform than others. The computation required to simulate each neuron at each time step in the simulation is generally fixed. The remaining time is then dedicated to processing the spikes received, the number of which depends on the how many neurons are sending spikes to the core and the activity of those connected neurons. This isn't known in advance in general, so some flexibility in the system with respect to the amount of computation available at each node is necessary to allow the application to work in different circumstances. Once this is known for a given network, the system could potentially be reconfigured with additional cores, allowing that network to be simulated in less time overall.

In this application, the assumption is made that there is some SpiNNaker application code that can process a number of neurons on each core.  This application will be set up with a series of synaptic matrices, one for each core that it can receive from.  This will enable the application to demultiplex the messages received from other cores, and direct them to the appropriate receiving neuron.  A more detailed description of the application code is found in \citep{Rhodes2018}.  In addition to the neuron application code, a Poisson spike generator code is also created, which will generate spikes randomly with a given rate using a Poisson process \citep{Gerstein1964}.  This is only expected to generate spikes rather than to receive them, though it might generate spikes for a number of source neurons.  The vertices can all record the spikes generated by each of the neurons, be it by Poisson process or neuron dynamics.

The setup for this application is as follows:
\begin{itemize}
  \item An application vertex is created for each group of simulated neurons (known as a Population), and another for each of the Poisson generators.
  \item A matching machine vertex is created for each of the above application vertices.  These machine vertices will be generated during the graph conversion, and each will be expected to take as a parameter the range of atoms assigned to it.
  \item An application edge is created to describe the connectivity between groups of neurons; the source vertex will be either a Poisson vertex or a neuron vertex but the target can only be another neuron vertex as described above.  This application edge will contain details of the neuron-to-neuron connectivity to allow the generation of the synaptic matrices.
  \item The neural network will be described as a set of neuron application vertices and Poisson source application vertices with neural application edges between them.
  \item Each machine vertex can generate data for the vertex it represents, including the parameters of the group of neurons, and, in the case of the neuron vertex, the source keys for each of the neurons it is to receive data from.
  \item Each machine vertex can tell the tools how many time steps it can run for given a space in SDRAM.  Note that the spike recording regions are sized assuming that every neuron spikes on every time step to ensure that the buffers don't overfill should this actually happen.
  \item Each machine vertex contains code to read spikes from the recorded data using the Buffer Manager.
\end{itemize}

Once the graph is built describing the neural network, the script will start execution of the graph, which will result in the execution of the simulation of the network.  This will first convert the application graph in to a machine graph by asking the application vertices how much resource they require for different ranges of neurons.  As with the Conway example, the tools will then go through the various stages until the execution of the simulation is complete.  Control will then return to the script and the user will be able to extract any recorded spikes and process these.

The live output example described in the Conway's use case also works similarly well in the neural networks use case.  Another extension more relevant here is the connection of an external device to the machine which will then be controlled by the network (e.g. a robotic device).  In this case, an extension of the virtual application vertex is made to represent the device.  When this is added to the application graph and the graph is executed, the tools will automatically detect this, and add a virtual chip to the discovered machine, with a placement constraint such that the device will be placed on this chip.  The tools will then operate as normal with edges to and from the device being routed as appropriate, and with the algorithms recognising that the chip is virtual and so making use of the adjacent chip when necessary.  Loading will recognize that the chip is virtual also, and so it won't attempt to load any data on to it.

More details of the neural network use case are described in \citep{Rhodes2018}, including other vertex types that are needed to simulate more complex networks.

\section{Future Work}\label{sec:future_work}
Although the SpiNNTools software provides many useful features, there are still many improvements that can be done.  Some of those include:
\begin{itemize}
  \item An application graph can only include application vertices.  Some utility vertices, such as the Live Packet Gatherer and the Reverse IP Tag Multicast Source may be more suited to being described as machine vertices.  To avoid having to provide both an application and machine vertex for those utilities, it might be better to allow an application graph to contain machine vertices, which are then simply copied to the machine graph during the conversion.
  \item It would be useful to provide some support at the C code level for demultiplexing messages from multiple atoms.  This would make writing code that supports application vertices easier.
  \item Data structures are currently manually written from the Python data generation and then read back from the C code.  This process could be streamlined by having a structure object in Python which can be used both to generate a C header file including the structure, and to write the data and to manage the SDRAM usage of the data in the Python code.
  \item The speed at which data is written to the machine is still quite slow.  A mechanism similar to that used to extract the data could be used to enhance this.
  \item Data extraction could be further improved by having cores on the Ethernet chip use the Ethernet adapter directly rather than having to go through SCAMP.
\end{itemize}

\section{Conclusions}\label{sec:conclusions}
We have described a software system SpiNNTools, which can be used to help to execute a problem described as a graph on SpiNNaker Neuromorphic machines.  We have described how the tools operate at the high level, by proceeding through a series of steps, which result in the mapping of the graph on to the machine, the execution of that graph and the obtaining of results from the execution.  We have then shown how this applies to two example problems, and the work that is still required to make each work on the machine.  We have shown that although SpiNNTools does not solve all the problems it abstracts away many of the steps required in the operation of the SpiNNaker machine, and therefore makes that operation easier.  Applications which make use of the tools will also see the benefits of any improvements without requiring large updates to their code-base.

\section*{Author Contributions}
This work presents the latest version of the SpiNNTools software package produced by the SpiNNaker group at the University of Manchester, UK. The current SpiNNaker software team is comprised of CB, DF, AG, OR and ABS, and led by AR, all of whom have made significant contributions to SpiNNTools. SD, DRL and LP are researchers within the SpiNNaker group, and worked on earlier versions of SpiNNaker software, and provided assistance with low-level programming and hardware interactions during performance analysis.  AR led the research and wrote the manuscript, while SBF leads the SpiNNaker project and supervised this work. All authors reviewed and refined the final manuscript.

\section*{Funding}
The design and construction of the SpiNNaker machine was supported by the EPSRC (UK Engineering and Physical Sciences Research Council) under grants EP/D07908X/1 and EP/G015740/1, in collaboration with the universities of Southampton, Cambridge and Sheffield and with industry partners ARM Ltd, Silistix Ltd and Thales. Ongoing development of the software is supported by the EU ICT Flagship Human Brain Project (H2020 785907), in collaboration with many university and industry partners across the EU and beyond, and our own exploration of the capabilities of the machine is supported by the European Research Council under the European Union’s Seventh Framework Programme (FP7/2007–2013)/ERC grant agreement 320689.

\section*{Acknowledgements}
SpiNNaker has been 15 years in conception and 10 years in construction, and many folk in Manchester and in our various collaborating groups around the world have contributed to get the project to its current state. We gratefully acknowledge all of these contributions.

\bibliographystyle{frontiersinSCNS_ENG_HUMS} 
\bibliography{pab_mendeley,library}

\end{document}